\documentclass[prb,twocolumn,msmath,superscriptaddress,amssymb,showpacs]{revtex4}

\usepackage{color} 
\usepackage{graphicx}
\usepackage{epstopdf}
\usepackage{bm}

\begin{document}

\DeclareGraphicsExtensions{.eps, .png, .jpg}
\bibliographystyle{prsty}

\title {Optical properties of V$_2$O$_3$ in its whole phase diagram}

\author{I. Lo Vecchio}
\affiliation{Dipartimento di Fisica, Universit\`{a} di Roma ``Sapienza", Piazzale A. Moro 2, I-00185 Roma, Italy}

\author{L. Baldassarre}
\affiliation{Center for Life Nano Science@Sapienza, Istituto Italiano di Tecnologia, V.le
Regina Elena 291, I-00186, Roma, Italy}

\author{F. D'Apuzzo}
\affiliation{Dipartimento di Fisica, Universit\`{a} di Roma ``Sapienza", Piazzale A. Moro 2, I-00185 Roma, Italy}
\affiliation{Center for Life Nano Science@Sapienza, Istituto Italiano di Tecnologia, V.le
Regina Elena 291, I-00186, Roma, Italy}

\author{O. Limaj}
\affiliation{Insititute of Bioengineering, \'Ecole polytechnique f\'ed\'erale de Lausanne (EPFL), 1015 Lausanne, Switzerland}

\author{D. Nicoletti}
\affiliation{Max Planck Institute for the Structure and Dynamics of Matter, Hamburg, Germany}

\author{A. Perucchi}
\affiliation{Sincrotrone Trieste, Area Science Park, I-34012 Basovizza, Trieste, Italy}

 \author{L. Fan}
\affiliation{National Synchrotron Radiation Laboratory, University of Science and Technology of China, Hefei, 230029, China}

\author{P. Metcalf}
\affiliation{Department of Physics, Purdue University, West Lafayette, Indiana 47907, USA}
 
\author{M. Marsi}
\affiliation{Laboratoire de Physique des Solides, CNRS-UMR 8502, Universit\'e Paris-Sud, F-91405 Orsay, France}

\author{S. Lupi}
\affiliation{CNR-IOM and Dipartimento di Fisica, Universit\`a di Roma ``Sapienza", Piazzale A. Moro 2, I-00185, Roma, Italy}
\date{\today}

\begin{abstract}
Vanadium sesquioxide V$_2$O$_3$ is considered a textbook example of Mott-Hubbard physics. In this paper we present an extended optical study of its whole temperature/doping phase diagram as obtained by doping the pure material with M=Cr or Ti atoms (V$_{1-x}$M$_x$)$_2$O$_3$. We reveal that its thermodynamically stable metallic and insulating phases, although macroscopically equivalent, show very different low-energy electrodynamics. The Cr and Ti doping drastically change both the antiferromagnetic gap and the paramagnetic metallic properties. A slight chromium content induces a mesoscopic electronic phase separation, while the pure compound is characterized by short-lived quasiparticles at high temperature. This study thus provides a new comprehensive scenario of the Mott-Hubbard physics in the prototype compound V$_2$O$_3$.
\end{abstract}
\maketitle

\section*{I. INTRODUCTION}
An electronic phase transition from a metallic to an insulating state exclusively driven by electron-electron interaction is called a Mott-Hubbard transition\cite{Mott-49, Hubbard-63}. This phenomenon can be found in some transition metal oxides having partially filled $d$ bands near the Fermi level. Although band theory would predict a metallic behavior, the Coulomb repulsion is so strong in these systems that it overcomes the kinetic energy and an insulating state, with one electron localized in each lattice site, becomes energetically favorable \cite{Gebhard, Imada-98}.

\begin{figure}[t]
\begin{center}
\leavevmode
\includegraphics [width=8cm]{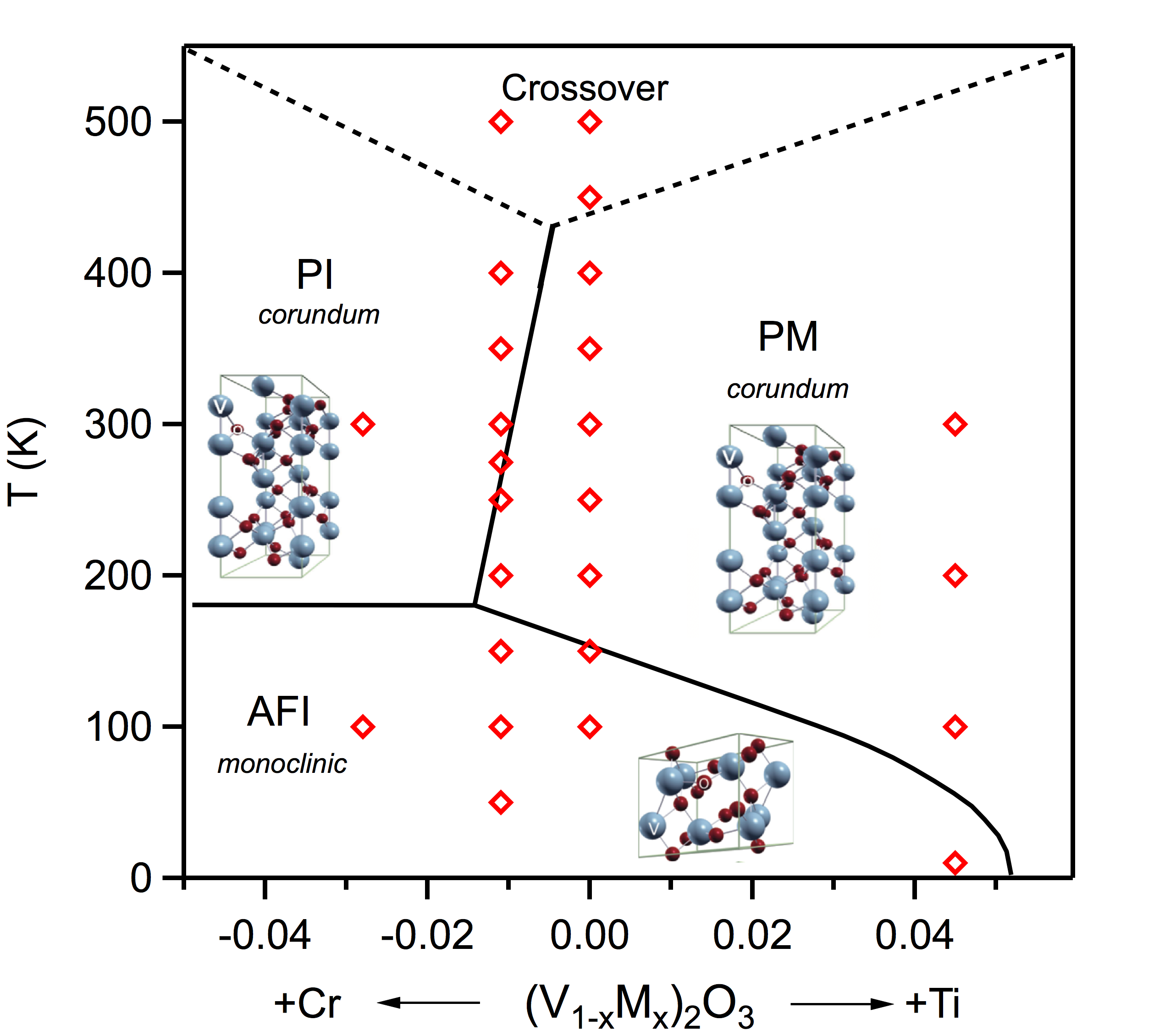}  
\end{center}
\caption{Temperature $vs$ doping phase diagram of V$_2$O$_3$ according to Ref. \onlinecite{McWhan-73}. In the PM and PI phase the compound has a corundum crystal structure, whereas the low temperature AFI phase shows a monoclinic symmetry (crystal structures have been adapted from Ref. \onlinecite{Hansmann-13}). Empty diamonds stand for temperature and doping values investigated in the present work.}
\label{Fig1}
\end{figure}

Vanadium sesquioxide V$_2$O$_3$, belonging to the vanadium oxide Magn\'eli phase \cite{Eyert-04, Baldassarre-07, Lo Vecchio-14, Postorino-07, Postorino-08, Perucchi-rev},
is considered, together with NiS$_2$ \cite{Perucchi-09}, a textbook example of the Mott-Hubbard physics. Since its discovery in 1960's \cite{Morin-59, McWhan-69, McWhan-73}, it has always been attracting a lot of interest because of its intriguing temperature-doping phase diagram \cite{Ding-14} shown in Fig. \ref{Fig1}.
The pure compound undergoes a first order metal-insulator transition (MIT) from a  paramagnetic metallic (PM) to a low-temperature antiferromagnetic insulating phase (AFI) around T$_N$=160 K with a seven orders of magnitude resistivity jump. This first order transition is accompanied by a crystal symmetry change from a corundum to a monoclinic structure. 
As soon as one slightly dopes the compound with chromium (from 0.5\% to 1.7\% Cr concentration) a new paramagnetic insulating phase (PI) appears, having the same corundum structure but slightly different lattice parameters \cite{McWhan-73}. The PM-PI transition is thus considered the prototype of the Mott-Hubbard transition, without any structural modification and/or magnetic ordering, but only driven by electron-electron interaction. The first order transition line presents a large temperature hysteresis ending at a second order critical point \cite{Limelette-03}. Above 400 K, a poor conducting regime is established, in which the single-particle spectral weight near the Fermi energy progressively disappears and the Drude peak is depleted in the low-energy optical conductivity \cite{Baldassarre-08}. Replacing vanadium with titanium, the N\'eel temperature $T_N$ monotonously decreases and above a 5\%Ti-content the AFI monoclinic phase is completely suppressed. The system becomes a metal with a corundum structure at all temperatures.

From the theoretical point of view, the mechanism that underlies the Mott-Hubbard MIT has longly been explained with a half-filled single band Hubbard model, in which the Coulomb parameter \textit{U}, i.e. the electron-electron repulsion, competes with the kinetic energy \textit{t}. Castellani \textit{et al.} \cite{Castellani-78} found a simple model in which the two 3\textit{d} electrons of the V$^{3+}$ ion are distributed into a singly degenerate occupied $a_{1g}$ and some partially filled doubly degenerate $e^\pi_g$ orbitals. These orbitals derive from a trigonal distorsion of the $t_{2g}$ manifold of the V 3\textit{d} states. Experimental evidence allows to treat this compound with a more realistic multi-orbital approach, which better describes its electronic properties \cite{Tomczak-09}. The XAS vanadium \textit{L}– edge study of Park \textit{et al.} \cite{Park-00} explored the phase diagram by varying both temperature and doping and summarized the respective ratio of $e^\pi_g$-$e^\pi_g$ to $a_{1g}$-$e^\pi_g$ occupations as: 1:1 (PM), 3:2 (PI), and 2:1 (AFI). After that, many-body models leading to a triplet state $S$=1 were formulated combining local density approximation (LDA) with the dynamical mean field theory (DMFT) \cite{Rozenberg-95, Georges-96}. LDA+DMFT calculations also succeeded in predicting a very important feature of V 3\textit{d} spectral function \cite{Held-01}, i.e. the quasiparticle (QP) peak at the Fermi energy in the PM phase, which was later observed by photoemission spectroscopy\cite{Mo-03, Mo-06, Marsi-09, Papalazarou-09, Panaccione_06}.

Despite intensive experimental and theoretical efforts \cite{Basov-11, Suga-11,Grieger-12, Kotliar-14}, many properties of this compound still remain elusive. For instance, a comparison between the metallic phase exhibited by pure, Cr and Ti-doped V$_2$O$_3$ is still missing. Furthermore, the low-energy electrodynamics in the AFI phase and the corresponding insulating gap has not been fully investigated yet as a function of increasing Cr doping. In general, a complete study of the evolution of the low-energy electrodynamics along the various electronic phases of V$_2$O$_3$ phase diagram has never been performed, to the best of our knowledge.
In this paper we present for the first time an extensive optical study of V$_2$O$_3$ over its whole phase diagram. In particular, we compare the optical conductivity data of differently doped compounds in the same phases. The paper is organized as follows: in Sec. II the experimental details are described. In Sec. III-A we present a general overview on the optical properties of all samples along their different phases. In particular from Sec. III-B to D we discuss the optical conductivity over the PM, PI and AFI phases. The summary and conclusion are finally presented in Sec. IV.

\section*{II. METHODS}
All samples are high quality (V$_{1-x}$M$_x$)$_2$O$_3$ single crystals grown at the Purdue University with various doping levels (x = 0; M = Ti, x=0.045; M = Cr, x = 0.011, 0.028). All of them were cut and polished along the $ab$ plane of the corundum structure.
We performed near-normal incidence reflectance R($\omega$) measurements using a Michelson interferometer in the frequency range from 50 cm$^{-1}$ to 20000 cm$^{-1}$ at temperatures between 10 and 500 K. 
A gold or silver (depending on the spectral range) coating was evaporated \textit{in situ} onto the sample surface and used as a reference. The optical conductivity was calculated by Kramers-Kronig transformations. Low-frequency reflectivity data were extrapolated with standard methods: Drude-Lorentz fit or constant lines.  
The reflectance of Shin et al. \cite{Shin-90}  was adapted to our data around 20000 cm$^{-1}$ in order to perform high-frequency extrapolation. Let us observe that R($\omega$) has a value of around 0.2 at high frequency for all crystals here measured. This means that the extrapolation with a unique curve should introduce a similar incertitude in the optical conductivity of various crystals.

\section*{III. RESULTS AND DISCUSSION}
\subsection*{A. Exploration of the optical response over the phase diagram}

\begin{figure*}[t]
\leavevmode
\centering
\begin{center}
\includegraphics[width=11.5cm]{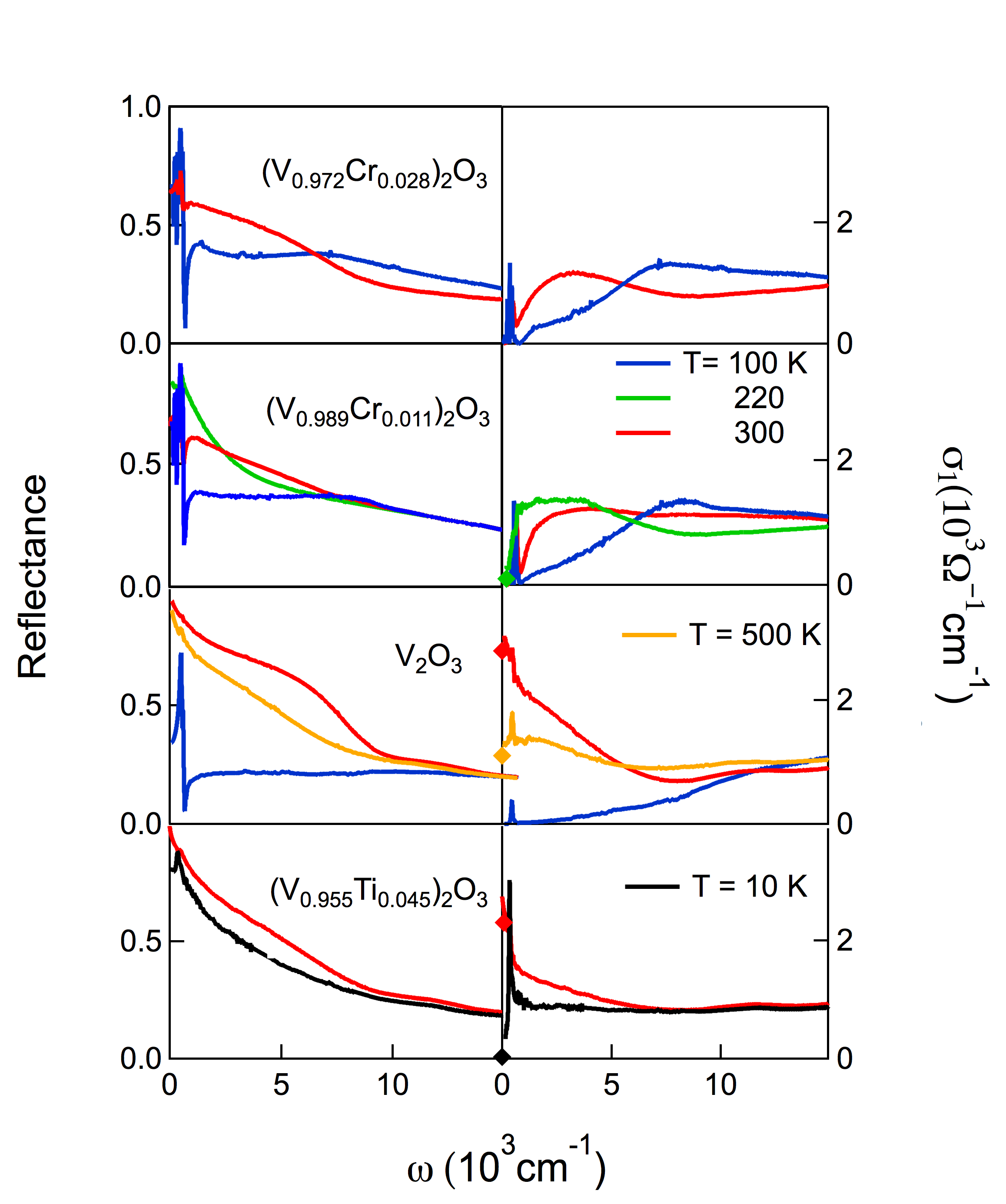}  
\end{center}
\caption{The near-normal incidence reflectance of four differently doped samples of V$_2$O$_3$ is reported on the left column at selected temperatures from far-infrared to visible frequencies. Right column shows the corresponding optical conductivities as obtained by Kramers-Kronig transformations on the same frequency scale and at the same temperatures of R($\omega$). Solid diamonds stand for $\sigma_{dc}$ values in the PM phase.}
\label{Fig2}
\end{figure*}

A general overview of the optical reflectance R($\omega$) of differently doped (V$_{1-x}$M$_x$)$_2$O$_3$  (M=Cr, Ti) samples in their various phases is shown in the left column of Fig. \ref{Fig2} for some representative temperatures and from 0 to 15000 cm$^{-1}$. The corresponding real part of the optical conductivity $\sigma_{1}(\omega)$ as obtained by Kramers-Kronig transformations is shown in the right column.
Reflectance spectra in the AFI phase are shown for all samples (blue curves) at 100 K apart from the Ti-doped one which, having a N\'eel temperature T$_N$$\simeq$60 K, is shown at 10 K (black curve). R($\omega$) of pure V$_2$O$_3$ is nearly flat up to about 8000 cm$^{-1}$ where an electronic absorption band starts to appear. In the Cr-doped samples (x = 0.011 and 0.028), R($\omega$) shows instead a small curvature below 5000 cm$^{-1}$ together with a weak electronic absorption at about 8000 cm$^{-1}$. Moreover, the AFI R($\omega$) value in pure V$_2$O$_3$ sets around 0.2, while it increases to about 0.4 in the Cr-doped samples. In the AFI phase of the Ti-doped system (x=0.045), the reflectance has a completely different frequency behavior. Indeed, R($\omega$) monotonously increases for $\omega \rightarrow0$ suggesting a strong reduction of the AFI electronic insulating gap.
At 300 K both the Cr-doped compounds are in their PI phase and show an appreciable enhancement of the reflectance in the mid-infrared. This clearly suggests a smaller insulating gap with respect to the AFI phase. 
Pure and titanium doped samples are in the PM phase at 300 K and, at this temperature, they show a metallic-like behavior with a reflectance that monotonously increases for $\omega \rightarrow0$. This behavior is similar to that of (V${_{0.989}}$Cr${_{0.011}}$)$_2$O$_3$ in its PM phase at 220 K. 

As explained in Sec. II, by performing Kramers-Kronig transformations, we obtained the real part of the optical conductivity $\sigma_{1}(\omega)$, plotted in the right column of Fig. \ref{Fig2}. In the AFI phase  $\sigma_{1}(\omega)$ shows, for all samples, a strong absorption peak around 500 cm$^{-1}$ which corresponds to the stretching phonon mode modulating the V-O distance.  Above the phonon absorption, $\sigma_{1}(\omega)$ shows a strong depletion (except for the Ti-doped system) which can be identified with the Mott-Hubbard gap. This gap reaches its maximum value in pure V$_2$O$_3$, is reduced to nearly half in Cr-doped samples and basically zero in the Ti-doped material. 

In the chromium doped compounds the PI conductivity is characterized by a nearly zero dc conductivity \cite{McWhan-73}, and an absorption band around 2500 cm$^{-1}$. For a 1.1\% content this band shifts to lower frequencies on crossing T$_{MIT}$ (green curve). This red-shift results, at low frequency, in a non-zero conductivity that can be well extrapolated to the dc value ($\sigma_{dc}$$\sim$250 $\Omega^{-1}$cm$^{-1}$), obtained by resistivity measurements on samples of the same batch. The absence of a clear Drude contribution in the PM phase of (V${_{0.989}}$Cr${_{0.011}}$)$_2$O$_3$ (see the difference between pure and 0.011 Cr-doped conductivities in Fig. 2) is due to a microscopic phase separation occurring in 0.011 Cr-doped system across the MIT that will be discussed in Sec. III-D.

In the pure compound, on crossing the N\'eel temperature, a filling of the Mott-Hubbard gap is induced (see also Section III-E). This filling results, below 1 eV, in a metallic absorption associated with the appearance of quasiparticles at $E_F$, in good agreement with what is observed in resistivity, specific heat \cite{McWhan-73}, and photoemission \cite{Mo-03} measurements. As the temperature increases, $\sigma_{1}(\omega)$ presents a strong temperature dependence, with a huge transfer of spectral weight from low to high frequency through an isosbestic point at about 6000 cm$^{-1}$.  For T=500 K (orange line) the pure sample is in the crossover region (see the phase diagram in Fig. 1), where the optical spectrum shows a low-frequency depletion in $\sigma_{1}(\omega)$, due to the progressive disappearance of quasiparticles at $E_F$ \cite{Baldassarre-08} (see Section III-E).

\subsection*{B. The PM phase}
According to well established band calculations, optical spectroscopy probes the V-V transitions up to 3.5 eV, while the V-O hopping occurs at higher energies \cite{Hansmann-13}. All the V 3\textit{d} optical absorption bands are explained in terms of transitions over a multi-orbital density of states\cite{Stewart-12}. The emerging Drude response when going from the AFI to the PM phase is then the result of a quasiparticle peak forming at the Fermi level at the expense of the Hubbard bands, separated by an energy U. In particular, both the $e^\pi_g$ and the $a_{1g}$ bands participate, in agreement with LDA+DMFT calculations \cite{Held-01}, to the formation of the QP peak at E$_F$ with different energy distribution and weight \cite{Rodolakis-10}. Together with the Drude term, transitions in the mid-infrared range are expected either from the lower Hubbard band to the quasiparticle peak, or from the QP peak to the upper Hubbard band. Due to the corundum crystal structure of V$_2$O$_3$, one might expect an anisotropic electrodynamic response. Specifically, a different Drude response along the $c$-axis and the $ab$-plane and, in principle, two absorption bands in the mid-infrared located at different characteristic energies.  
Although most of transport measurements in the PM phase have been performed along the $ab$ hexagonal plane, the very few $c$-axis dc transport data present in literature show \cite{Feinleib-67} an anisotropy on the order of 5$\%$, with the $c$-axis more conductive than the $ab$-plane.
We remind that the whole optical data in this paper and the following discussion refer to the hexagonal $ab$ plane.

The PM optical conductivity for the pure and Ti-doped crystals are shown in Fig. 3 (upper panels), at 200 K and 300 K. In both systems one can  observe a clear Drude tail below 1000 cm$^{-1}$ and a broad peak in the mid-infrared (MIR band). A high-frequency absorption reminiscent of an incoherent Hubbard transition, which will be discussed more properly in Section E, is still present above 10000 cm$^{-1}$. As $\sigma_{1}(\omega)$ in both crystals has been measured along the $ab$ hexagonal plane, these spectral features should be mainly related to the $e^\pi_g$ orbitals, which are oriented along this plane. 
The whole infrared conductivity for both systems increases by reducing T through a transfer of spectral weight from the high-frequency incoherent transitions. 
This indicates that coherent transport develops at low temperature, yielding a better metallic state.

\begin{figure}[h]
\begin{center}
\leavevmode
\includegraphics [width=9.5cm]{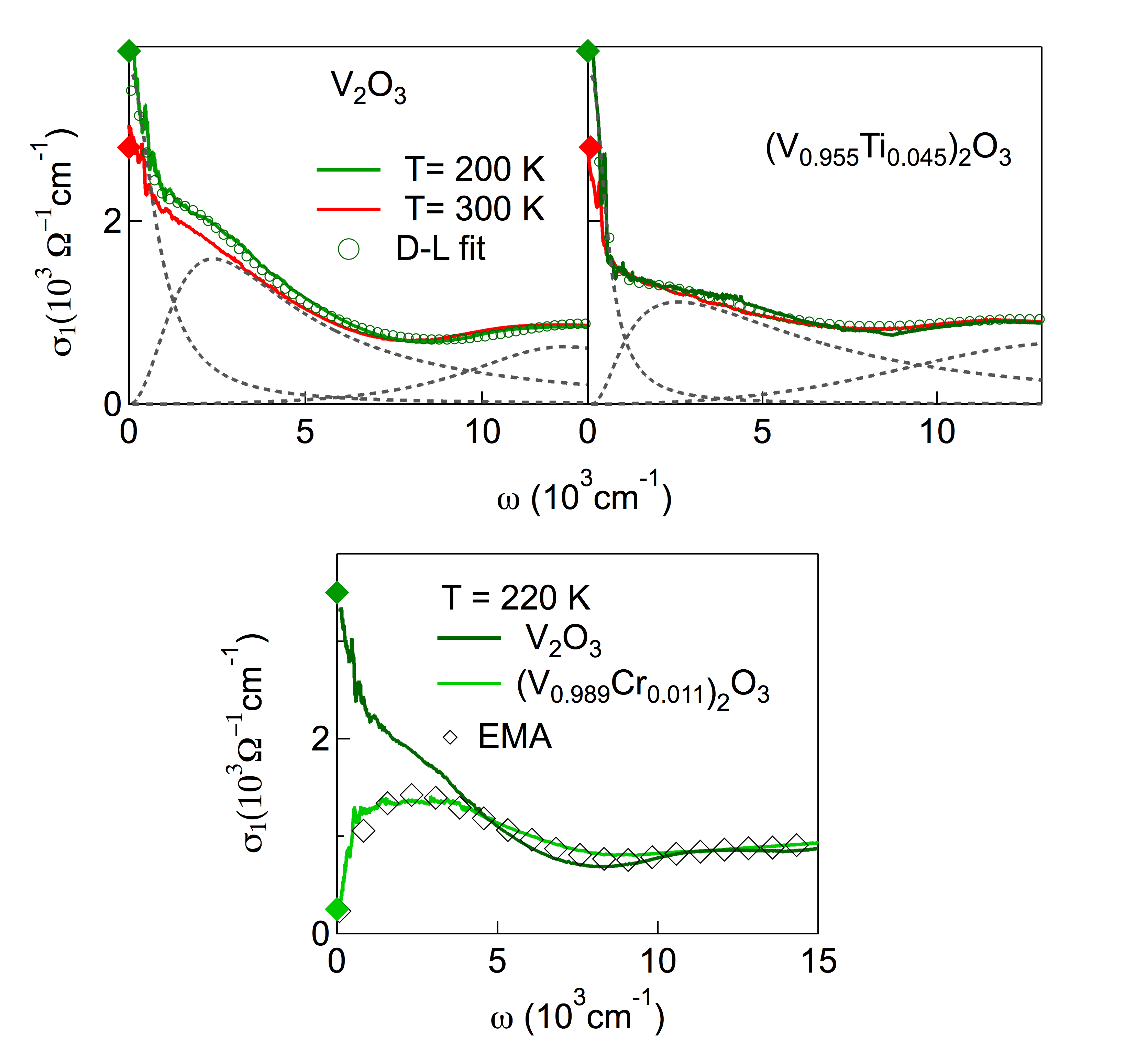}  
\end{center}
\caption{Top panels: The real part of the optical conductivity in the PM phase at 300 K (red line) and 200 K (green line) is shown for V$_2$O$_3$ and (V$_{0.955}$Ti$_{0.045}$)$_2$O$_3$, respectively. A Drude-Lorentz fit (empty circles) is also shown for T= 200 K along with the different spectral components (dashed lines): a Drude term, a MIR band, and the high-frequency Hubbard transitions. Bottom panel: Comparison of the optical conductivities of pure V$_2$O$_3$ and of the 1.1\% chromium doped sample in their PM metallic phase at T= 220 K. The poor metallic behavior of the Cr-doped system is due to an electronic phase separation in a large temperature range across the Mott-Hubbard transition \cite{Lupi-10} which can be described in terms of an effective medium approximation EMA (see text).Solid diamonds stand for $\sigma_{dc}$ values.}
\label{Fig3}
\end{figure}

In order to compare more quantitatively the PM electrodynamics of both crystals, we have fitted $\sigma_{1}(\omega)$ at 200 K through a multi-component Drude-Lorentz (D-L) model including the Drude term, and three Lorentz peaks: the stretching phonon around 500 cm$^{-1}$, a MIR band and the Hubbard band. The complex optical conductivity is written as:
\begin{equation}
\tilde{\sigma}(\omega)=\frac{\omega_p^2\tau}{4\pi(1-i\omega\tau)}+\frac{\omega}{4\pi i}\sum_j\frac{S_j^2}{\omega_j^2-\omega^2-i\omega\gamma_j}
\label{DrudeLorentz}
\end{equation}
In the Drude term $\omega_p$ is the plasma frequency and $\tau$ is the scattering time. 
The Lorentz bands are peaked at finite frequencies $\omega_j$ with strength $S_j$ and width $\gamma_j$. The whole fitting curve and the single components (except for the phonon), are represented in Fig.\ref{Fig3} (empty circles and dashed lines, respectively), for both crystals at 200 K.

\begin{table}[h]
\begin{tabular}{|c|c|c|}
\hline  & x=0 & x(Ti)=0.045 \\ 
\hline $\omega_P$ & 10800 & 12300 \\ 
\hline $\Gamma$=1/$\tau$ & 530 & 500 \\ 
\hline S$_{MIR}$ & 23500 & 20200 \\
\hline $\gamma_{MIR}$ & 5500 & 6800 \\
\hline $\omega_{MIR}$ & 2350 & 2350 \\
\hline 
\end{tabular} \\
\caption{Drude-Lorentz fit parameters in cm$^{-1}$ for T=200K.}
\end{table}

From the fitting results one can estimate the ratio of the in-gap spectral weight (Drude+MIR band) between pure and Ti-doped crystals. This ratio (which is basically indipendent of temperature)
is nearly 1.1 indicating that a slight reduction of metallicity is induced by Ti-substitution. As V presents a $3d^{3}$$4s^{2}$ electronic configuration while Ti has only two electrons in the 3$d$ orbitals ($3d^{2}$$4s^{2}$), this suggests that titanium acts like an electronic acceptor adding x(Ti) holes in the crystal. Starting from the formula unit (V$_{1-x}$Ti$_x$)$_2$O$_3$ this corresponds to an electron density ratio 1.05  for x(Ti)=0.045, in good agreement with the experimental value 1.1 determined above. 

From the fitting parameters reported in Table I one can also calculate for each sample the spectral weight ratio \cite{DeGiorgi-11,Baldassarre-12}:

\begin{equation}
SW_{D}/SW_{in-gap}=\frac{\omega_p^2}{\omega_p^2+S_{MIR}^2}
\label{Eq3}
\end{equation}

\noindent Where $SW_{D}$ ($\omega_p^2$) is the intensity of the Drude component while $SW_{in-gap}$ also contains the MIR contribution to the optical conductivity. This quantity, obtained from the experimental data, provides an estimate of the degree of electronic correlation of any material \cite{Qazilbash-09, DeGiorgi-11,Baldassarre-12}. When $SW_{D}/SW_{in-gap}$ is close to zero this means that the electron-electron correlations are strongly renormalizing the coherent spectral weight; for $SW_{D}/SW_{in-gap}$$\to$1, one approaches a conventional metallic case like gold and silver \cite{Qazilbash-09}. 
At 200 K, $SW_{D}/SW_{in gap}$$\sim$0.18 for the pure sample. This indicates that V$_2$O$_3$ is a strongly correlated metal, where electron-electron repulsion hugely renormalizes the quasiparticle spectral weight in favor of the mid-IR incoherent part. In V$_2$O$_3$ films \cite{Stewart-12} containing both contributions from the $ab$-plane and $c$-axis, the same ratio at 200 K is 0.12 in quite good agreement with our estimate. In the Ti-doped system $SW_{D}/SW_{in-gap}$$\sim$0.28 pointing out that although correlations still play a strong role, their effect is reduced with respect to pure V$_2$O$_3$. This last result can be better understood in terms of Ti-doping effect previously discussed. Indeed, Ti acting like an acceptor, displaces V$_2$O$_3$ from the half filling state and this naturally implies that the effects of electronic correlations are reduced.

It is also worth noticing that the PM optical conductivity of (V$_{1-x}$Cr$_x$)$_2$O$_3$ with x(Cr)=0.011 (see Fig. 3, bottom panel) does not show a metallic behavior. Indeed its PM conductivity at 220 K shows a strong MIR absorption looking very similar to that of pure V$_2$O$_3$ material, without the presence of a low-frequency Drude term. This drastic depletion of the Drude is due to the proximity with the Mott transition (see the introduction section), and in the next Section we will see that a mesoscopic electronic phase separation strongly influences the properties of this sample.

\subsection*{C. Phase coexistence at the Mott transition}
\begin{figure*}[t]
\begin{center}
\leavevmode
\includegraphics [width=14.5cm]{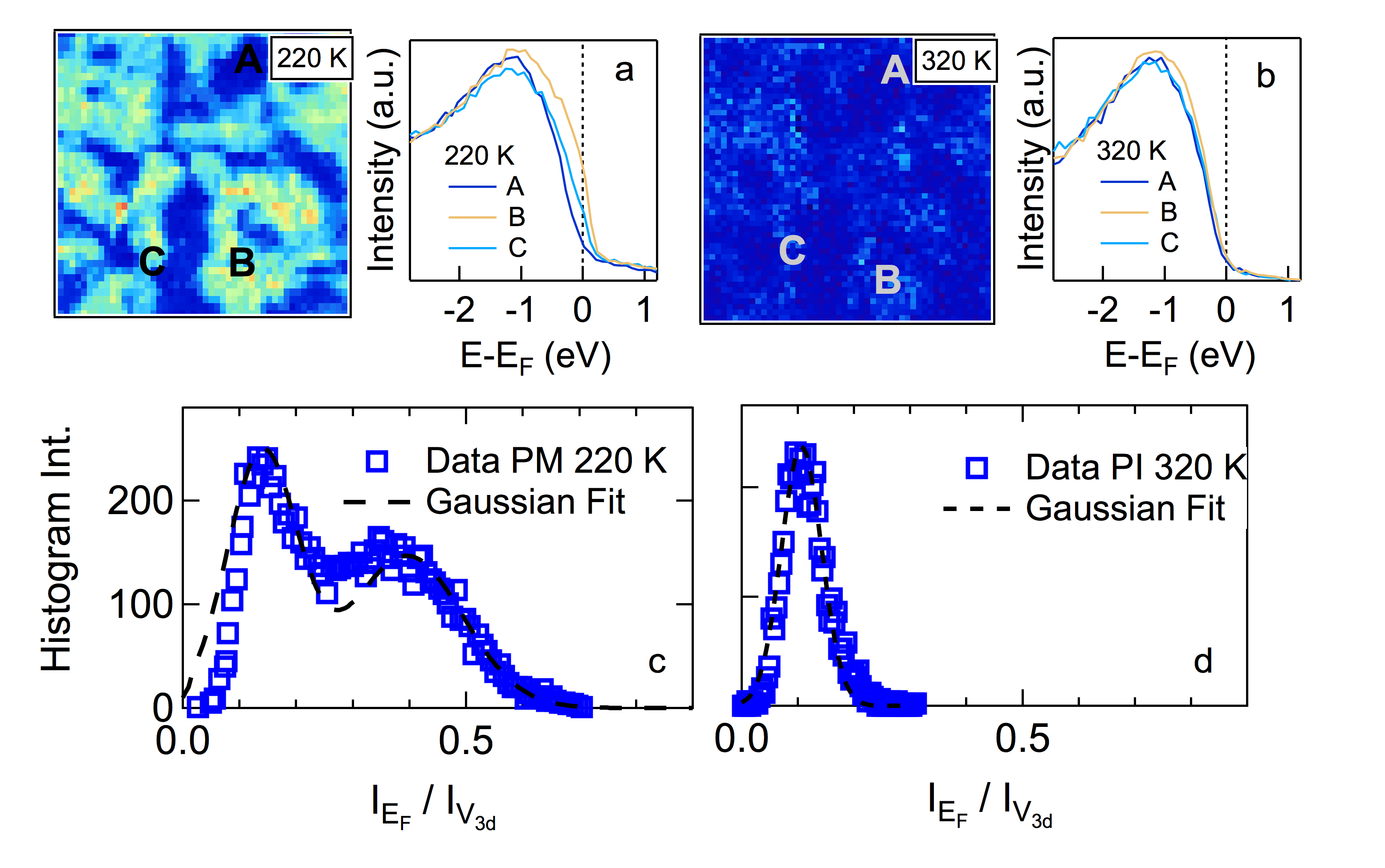}  
\end{center}
\caption{(a)(b): Scanning photoemission microscopy images and photoemission spectra on (V$_{0.989}$Cr$_{0.011}$)$_2$O$_3$, collected at 27 eV photon energy and at different temperatures on a 50 $\mu$m by 50 $\mu$m sample area. The pictorial contrast between metallic (yellow) and insulating (blue) zones is obtained from the photoemission intensity at the Fermi level \cite{Gunther-97}, normalized by the intensity on the V3d band (binding energy -1.2 eV). Inhomogeneous properties are found within the PM phase at T=220 K, while at 320 K in the PI phase a homogeneous insulating state is obtained. (c, d)) Histogram of the intensity distribution of the PM (PI) photoemission map (open symbols) as extracted from the map at 220 (320) K.}
\label{Fig4}
\end{figure*}

In this section we focus on the first order transition from a paramagnetic insulating to a paramagnetic metallic phase that happens in doped compounds with a chromium percentage varying from 0.5\% to 1.7\%. In particular, the MIT in the 1.1\% sample occurs at $\sim$ 260 K upon cooling; upon heating thereafter the sample, the PM phase remains stable up to 275 K, and the PI phase is fully recovered only for temperatures above 300 K. The optical conductivity, for light polarization along the $ab$ plane at 220 K, is shown in the bottom panel of Fig. 3. In contrast with the pure compound in the same macroscopic PM phase (see the same panel), the doped compound shows a depletion of the Drude response and a lower dc value:  $\sigma_{dc}(x=0.011)$$\sim$250 $\Omega^{-1}$cm$^{-1}$ to be compared to  $\sigma_{dc}(x=0)$$\sim$3500 $\Omega^{-1}$cm$^{-1}$.

The absence of a metallic contribution in the optical spectra of the 1.1\% sample is at variance with photoemission measurements recently achieved on crystals from the same batch. Indeed, photoemission data show, when performed with sufficient bulk sensitivity, a clear metallic quasiparticle peak \cite{Mo-06,Marsi-09}. In order to clarify these apparently contradictory results, we performed a Scanning Photoemission Microscopy (SPEM) experiment with sub-micrometric spatial resolution. Photons at 27 eV were focused through a Schwarzschild objective, to obtain a submicron size spot. The measurements were performed in ultra-high vacuum (P$<2$$\times10^{-10}$ mbar) on well oriented $ab$ surfaces of (V$_{0.989}$Cr$_{0.011}$)$_2$O$_3$ prepared \textit{in situ}. Photoelectrons were collected at an emission angle of 65 degrees from the normal, which makes the photoemission spectra and images surface sensitive. A standard procedure was used to remove topographic features
from the images presented in Fig. \ref{Fig4}a,b, and to obtain an unambiguous contrast between metallic and insulating regions \cite{Sarma-04}. More specifically, the spectral intensity at the Fermi level E$_F$, which is high in the PM phase and absent in the PI phase, was divided by the V 3\textit{d}-band intensity. As this latter is essentially constant in the PI and PM phases, two-dimensional maps of this ratio provide a genuine representation of the lateral variation of metallic and insulating domains. 

The sub-micrometric maps of the photoelectron yield at $E_F$ \cite{Lupi-10} are plotted in Fig. \ref{Fig4}a,b together with their representative photoemission spectra at 220 K (PM phase) and 320 K (PI phase).  It should be pointed out that our experimental geometry \cite{Dudin} and photon energy (27 eV) did not allow the collection of sufficiently bulk sensitive photoelectron spectra. Thus the spectra corresponding to the metallic phase do not show an evident quasiparticle peak. Nevertheless, the difference between insulating and metallic phases is unambiguous. 
Within these maps we can resolve metallic (yellow zones), and insulating domains (blue zones), which appear for $T<T_{MIT}$ and persist in the PM phase. A uniform insulating behavior is found in the PI phase, showing that the PM state only is strongly phase separated. Moreover, when the sample is further cooled down, the metallic and insulating domains are still found in the same sample position and with a similar shape \cite{Lupi-10}. This memory effect clearly shows that mesoscopic domains form around specific points in the sample, for instance around dopant Cr ions \cite{Grieger-14}. These nucleation centers can be related to structural inhomogeneites in the material and then have an extrinsic origin. However, the tendency towards the electronic phase separation is an intrinsic phenomenon associated with the correlation energy providing the thermodynamic instability in the vicinity of the Mott transition.

To the electronic phase separation in the PM phase, as observed by SPEM, it also corresponds a structural phase separation. In fact, the structure of (V$_{0.989}$Cr$_{0.011}$)$_2$O$_3$  in the PM phase has been described in terms of a coexistence of an $\alpha$-phase, with lattice parameters close to those of metallic V$_2$O$_3$ at 400 K, and a $\beta$-phase with basically the same structure as at higher  Cr content \cite{Robinson-75,McWhan-73}.
SPEM and structural data therefore provide a much deeper understanding of the poor metallic behavior in the PM phase of (V$_{0.989}$Cr$_{0.011}$)$_2$O$_3$, as shown by the pseudo-gap low-frequency behavior of the optical conductivity (see Fig. \ref{Fig3}). In an inhomogeneous (two-phases) system investigated with an electromagnetic field having wavelengths larger than the mesoscopic domain sizes, the dielectric properties can be described through an effective medium approximation (EMA). In this case the complex dielectric function $\epsilon_{EMA}$ can be calculated utilizing the optical response of the two components $\epsilon_{I}$ (insulating part) and $\epsilon_{M}$ (metallic part), weighted by the respective volume fraction \textit{f} and depolarization factor \textit{q} \cite{Lupi-10}: 

\begin{equation}
f\frac{\epsilon_{I}-\epsilon_{EMA}}{\epsilon_{I}+\frac{1-q}{q}\epsilon_{EMA}}+(1-f)\frac{\epsilon_{M}-\epsilon_{EMA}}{\epsilon_{M}+\frac{1-q}{q}\epsilon_{EMA}}=0
\end{equation}

In particular, the volume fraction $f$ can be determined from the SPEM images of the spectral weight at E$_{F}$ (Fig. \ref{Fig4}a). An example of the intensity distribution histogram at E$_{F}$, extracted from the PM maps is shown at 220 K in Fig. \ref{Fig4}c.  Here, the bimodal distribution clearly indicates two main phases present in the sample. An insulating phase, corresponding to the blue color in the photoemission map, and a metallic phase corresponding to the yellow color. The broad minimum between these two components corresponds to areas where metallic and insulating regions are too small to be resolved with our spatial resolution. Furthermore, the PM histogram can be fitted by a sum of two Gaussian curves (dashed line), whose relative intensities provide the insulating and metallic volume fractions. The fit gives $f_M$=0.45$\pm$0.05 ($f_I$=0.55$\pm$0.05) for the metallic (insulating) component ($f_M$+$f_I$=1).
Remarkably, at 320 K one Gaussian (corresponding to the insulating phase) is sufficient to fit the intensity distribution histogram (Fig. \ref{Fig4}d), as expected from the corresponding photoemission map, which shows an almost homogeneous insulating distribution.  

The EMA reconstruction can be made at any T, by using as input parameters of Eq. 3 the metallic and insulating dielectric functions as measured in V$_2$O$_3$ at 400 K \cite{Baldassarre-08} and (V$_{0.972}$Cr$_{0.028}$)$_2$O$_3$, respectively, estimating the volume fraction of metallic and insulating components from the SPEM images and leaving $q$ as the only free-fitting parameter.
An example of EMA reconstruction at T=220 K is shown in Fig. 3 (bottom panel, empty squares) where $q$ is found to be 0.30$\pm$0.05 at 220 K (0.40$\pm$0.05 at 300 K), similar to the value achieved in VO$_2$ across its first order MIT \cite{Qazilbash-07}. One can remark that the quality of reconstruction scarcely depends on $q$ since $f\!>\!q$. Moreover, in this limit, the MIT transition has a percolative nature, thus explaining the low dc conductivity ($\sigma_{dc}$$\sim$250 $\Omega^{-1}$cm$^{-1}$) value measured in samples coming from the same batch. 
Therefore, the mesoscopic electronic phase separation around $T_{MIT}$ determines the optical and $dc$ behavior of (V$_{0.989}$Cr$_{0.011}$)$_2$O$_3$ indicating that the PM phase of this specific sample (and probably of all doped samples in the 0.5\% to 1.7\% Cr interval, see the phase diagram in Fig. 1), is completely different from those of V$_2$O$_3$ and  (V$_{0.955}$Ti$_{0.045}$)$_2$O$_3$, where a phase separation has not been observed \cite{Mansart-12}.

\subsection*{D. Crossover region}
In the high temperature crossover region of the phase diagram (Fig. 1), a poor metal conducting regime has been observed through $dc$ transport measurements \cite{McWhan-73} and this has been associated to the progressive disappearance of quasiparticle coherence \cite{Mo-06}. In this Section we will discuss the behavior of the optical conductivity of the pure V$_2$O$_3$ sample when approaching the crossover region. 

In V$_2$O$_3$ QPs develop for T$>$T$_N$. This can be observed from the optical conductivity which, for sake of clarity, is shown in the inset of Fig. 5a at selected temperatures covering the AFI, PM and crossover region phases. In the AFI phase $\sigma_{1}(\omega)$ shows a strong depletion in the mid- and near-infrared followed by an increasing absorption associated with the Mott-Hubbard gap (see also Section III-E). The MIT at T$_N$$\simeq$160 K corresponds to an abrupt filling of the gap due to a huge transfer of spectral weight at low-frequency, and the highest conductivity is achieved just above the MIT. By raising the temperature towards the crossover region, the transfer of SW is reversed: the far- and mid-infrared conductivities start to be depleted in favor of the near-infrared and visible conductivity. In particular, for $T<$400 K, $\sigma_1(\omega)$ has an increasing behavior $vs.$ $\omega$ indicating the presence of a low-frequency QP coherent contribution which can be well fitted through a Drude term (see Section III-B). Above this temperature $\sigma_1(\omega)$ shows a slope change in the far-infrared (it decreases for $\omega\rightarrow$0), which suggests the disapppearence of coherent QP excitations in agreement with the enhancement of the dc resistivity in entering the crossover regime \cite{McWhan-73}.

\begin{figure}[t]
\begin{center}
\leavevmode
\includegraphics [width=8.5cm]{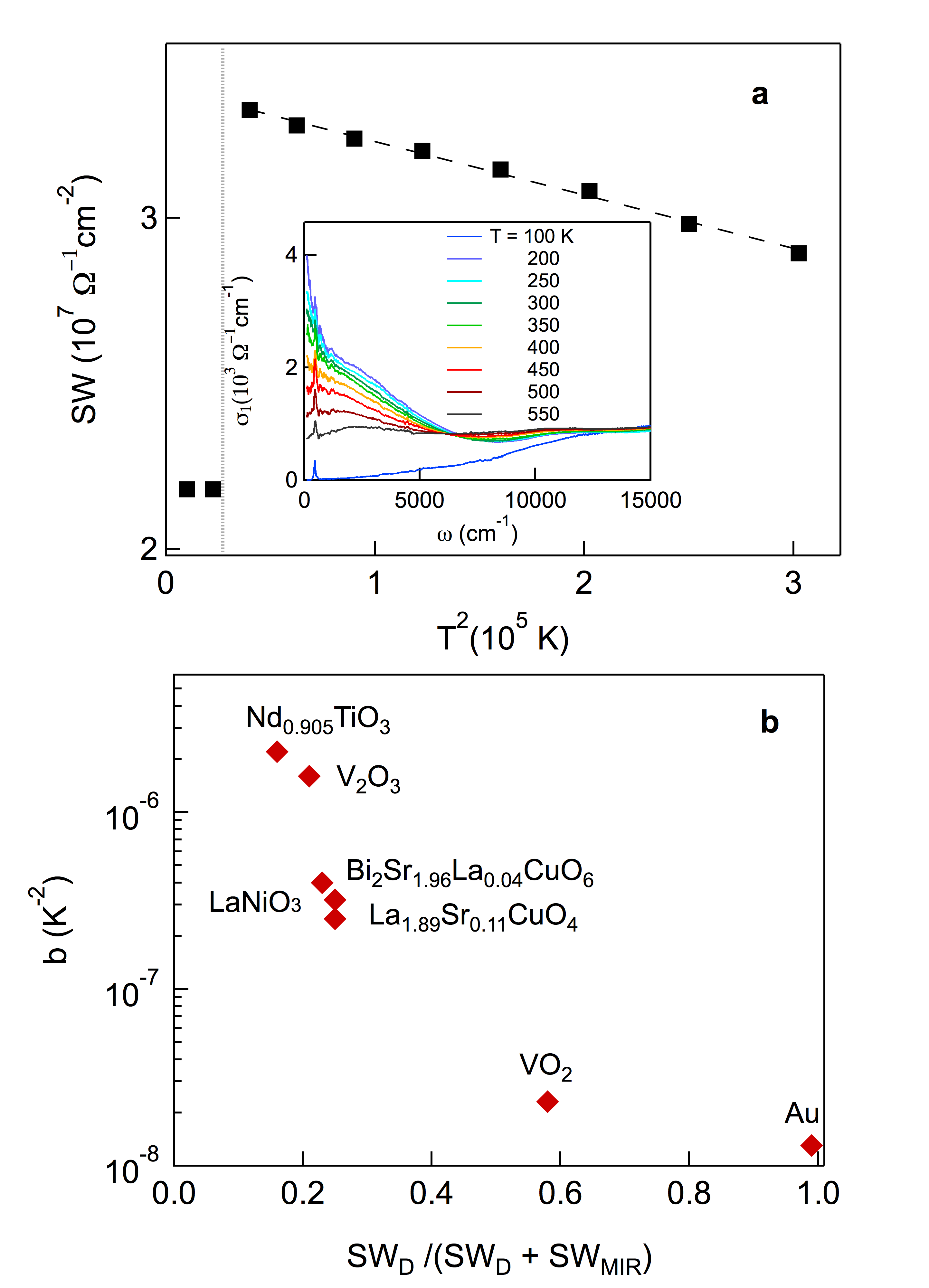}  
\end{center}
\caption{(a) Spectral weight of pure V$_2$O$_3$ integrated up to 8000 cm$^{-1}$ and plotted as a function of T$^2$, showing a linear behavior. The optical conductivity from 100 K to 550 K is shown in the inset. (b) Correlation diagram between two different ways of estimating the strenght of electron-electron interactions in various materials.}
\label{Fig5}
\end{figure}

To quantify the reduction of coherence for increasing T, we plot in Fig. \ref{Fig5}a the in-gap optical spectral weigth SW$_{in-gap}$ (SW$_{in-gap}$=SW$_D$+SW$_{MIR}$). A similar value of the SW can be achieved by integrating the optical conductivity up to a cut-off frequency $\Omega$=8000 cm$^{-1}$ corresponding to the minimum (i.e. the pseudo-plasma edge $\Omega_p$) in $\sigma_1(\omega)$ at 200 K. 
As a function of T (see Fig. \ref{Fig5}a), SW$_{in-gap}$ is small and nearly constant in the AFI state and shows a large, discontinuous jump on crossing the MIT presenting, just above T$_N$, its maximum value. This indicates, quantitatively, the birth of quasiparticles in the metallic state. However, for a further increase of T, a progressive loss of the spectral weight starts to be evident. This loss mainly reflects the lowering of the metallic contribution, and the access to an incoherent state.

The temperature dependence of SW$_{in-gap}$ nicely follows the quadratic relation\cite{Toschi-05} $SW_{in-gap}(T)=W_0-BT^2$ (dashed line in Fig. 5a), where $B$ describes the thermal response of carriers. The $T^2$ dependence of $SW_{in-gap}(T)$ is a general feature, shared by conventional and correlated metals. This dependence can be explained already in a tight-binding approach through the Sommerfeld expansion of the kinetic energy, which is proportional to the optical spectral weight.  While in conventional metals a nearly complete spectral weigth recovering at any T is achieved if $\Omega\sim\Omega_p$, resulting in a vanishingly small $B(\Omega_p)$, electronic correlations strongly renormalize both $B$ and $W_0$. Therefore, the parameter $b=$B($\Omega_p)$/$W_{0}$($\Omega_p$) can be used to compare the degree of electronic correlation in different systems \cite{Baldassarre-08, Nicoletti-10}.
From the fit of SW $vs.$ T in Fig. 5a we estimate $b=1.6\times$10$^{-6}$K$^{-2}$. This value is comparable to that of Nd$_{0.905}$TiO$_3$\cite{Timusk-06} (2.2$\times$10$^{-6}$K$^{-2}$), sizably larger (nearly a factor 100) than a conventional metal like gold\cite{Ortolani-05} (1.3$\times$10$^{-8}$K$^{-2}$) and VO$_2$ (this paper, 2.5$\times$10$^{-8}$K$^{-2}$) and also larger than other correlated materials like La$_{1.89}$Sr$_{0.11}$CuO$_4$ \cite{Ortolani-05}
(4.5 $\times$10$^{-7}$K$^{-2}$), Bi$_2$Sr$_{1.96}$La$_{0.04}$CuO$_6$ \cite{Nicoletti-11}
(2.5$\times$10$^{-7}$K$^{-2}$), LaNiO$_3$ (this paper, 3.2$\times$10$^{-7}$K$^{-2}$).

As discussed in Section III-B, the degree of electronic correlation is currently estimated in literature through the ratio $SW_{D}/SW_{in-gap}$ \cite{Qazilbash-09}, where $SW_{D}$ is the spectral weight of the Drude component while $SW_{in-gap}$ also contains the MIR contribution to the optical conductivity. Indeed, electronic correlations reduce, at a fixed T, the coherence of QPs transferring part of their SW from the Drude term to a MIR band. Moreover,  electronic correlations also increase the temperature dependence of SW with respect to a non-correlated system.  
Therefore, it is interesting to compare these two aspects of electronic correlations in different materials. In Fig. 5b we represent a correlation diagram between the parameter $b$ and $SW_{D}/SW_{in-gap}$ for V$_2$O$_3$ and the other materials previously discussed. A strong reduction of the Drude coherent spectral weight in favor of the incoherent MIR band, i.e. a low value of the $SW_{D}/SW_{in-gap}$ ratio clearly corresponds to a strong transfer of SW vs.T as indicated by a large value of $b$. In V$_2$O$_3$ $SW_{D}/SW_{in-gap}$ is minimized and $b$ is maximized suggesting that very strong correlations play a fundamental role in any temperature region of the phase diagram. A similar behavior can be observed in Nd$_{0.905}$TiO$_3$, where the effect of correlations are even more pronounced. 

On the other hand, the position of VO$_2$ in the correlation diagram suggests that electronic correlations do not play a strong role in both the temperature dependence of the spectral weigth and in the coherent/incoherent spectral weigth ratio. Therefore, it is hard to say that the main driven force of the MIT in VO$_2$ is the Hubbard mechanism.
Finally, the absence of electronic correlations in gold reflects in a very weak T dependence of the SW, $i.e.$ in a very small $b$ value, and in an optical conductivity that can be described in terms of a pure Drude component.

\subsection*{E. The AFI phase}

\begin{figure}[t]
\begin{center}
\leavevmode
\includegraphics [width=8.5cm]{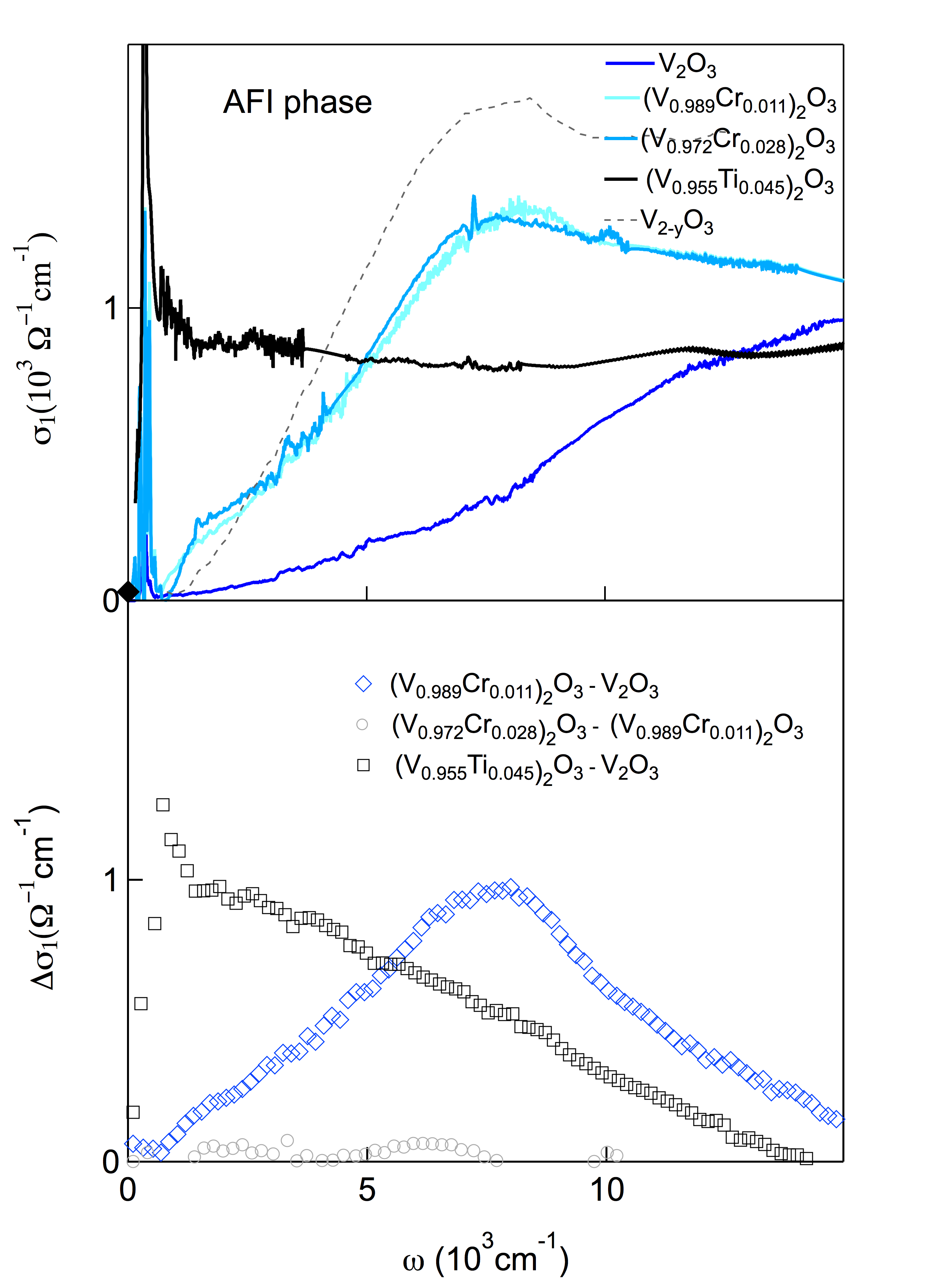}  
\end{center}
\caption{Upper panel: Optical conductivity of V$_2$O$_3$ in its antiferromagnetic insulating phase for various dopings. The dashed curve corresponds to a oxygen-vacant crystal (Ref. \onlinecite{Rozenberg-95}). Bottom panel: A strong reduction of the AFI optical gap can be observed in both Cr- and Ti-doping as highlighted by subtracting the pure sample curve from the Cr- and Ti-data.}
\label{Fig6}
\end{figure}

The AFI optical conductivities at 100 K for pure and Cr-doped systems and at 10 K for the Ti-doped crystal are plotted in the upper panel of Fig. \ref{Fig6}.  
An optical gap can be observed in pure V$_2$O$_3$ \cite{Thomas-94}. A slight Cr-doping (1.1 \%) unexpectedly generates a strong collapse of the gap which nearly reduces by a factor of two. Moreover, a broad peak appears around 8000 cm$^{-1}$ extending smoothly at higher energy. This peak can be better highlighted by subtracting the pure sample curve from the 1.1 \% data as shown in the lower panel of Fig. \ref{Fig6}. An increasing Cr-doping does not further change the AFI gap as observed by the difference $\sigma_{1}(\omega, 0.028)$-$\sigma_{1}(\omega, 0.011)$. Such a strong reduction of the AFI gap $vs.$ the Cr content is at variance with the smooth variation of the N\'eel temperature which sligthly increases from 160 K for x=0 to nearly 180 K for any Cr-doping.
A similar gap collapse has also been measured by Rozenberg \textit{et al.}\cite{Rozenberg-95} in a V$_{2-y}$O$_3$ single crystal with y=1.3\% of vanadium vacancies (dashed line in the upper panel of Fig. \ref{Fig6}) where T$_{N}$ reduces to $\simeq$50 K.

The effect of Ti-doping is even more robust. The AFI gap, well visible in pure and Cr-doped materials basically disappears in favor of a flat electronic background which extends from the mid-IR to the visible range. Only in the far-IR one can observe a small gap on the order of  300 cm$^{-1}$ which smoothly interpolates to a dc conductivity of about 30 $\Omega^{-1}$cm$^{-1}$ (solid black diamond, see Fig. 6). This mirrors a huge transfer of spectral weight towards low-frequency as indicated by the difference $\sigma_{1}(\omega, 0.045)$-$\sigma_{1}(\omega, 0)$. Although little is known about the AF order in the Ti-doping region, at the actual doping level (x(Ti)=0.045) T$_N$ interestingly has a still finite value of about 60 K, comparable to that of V$_{2-y}$O$_3$ (y=1.3\%). 

Although a theoretical calculation would be necessary to quantitatively explain all features related to the AFI transition in the whole phase diagram, the behavior of the optical conductivity can be qualitatively understood from the effects of Cr and Ti incorporation in the lattice in a general LDA perspective.
Indeed, the AFI phase is characterized by a monoclinic I2/a symmetry, whose lattice parameters are $a_{m}$ = 7.255 \AA, $b_{m}$ = 5.002 \AA, $c_{m}$ = 5.548 \AA, and a monoclinic angle $\beta$= 96.752 degrees. As a consequence, both the vertical and the horizontal V-V bond lengths increase determining a finite increase in the lattice volume compared to the corundum structure. The monoclinic bond lengths are basically constant over the whole phase diagram, suggesting that the different behavior of the optical conductivity across the transition should be deduced from the different properties of the metallic/insulating phase above T$_N$. 

Ti substitution plays a double role. The incorporation of Ti into V$_2$O$_3$ results in a compression of the corundum lattice \cite{Kuwamoto-76}. This suggests a reduction of the $U/W$ ratio $i.e.$ an increase of the density of states at the Fermi energy. This is also in agreement with the fact that Ti$_2$O$_3$ is metallic at all temperatures. However, a larger density of states at $E_F$ is partially reduced by the fact that Ti acts like an acceptor (see discussion in Section III.B), thus determining a smaller electron density with respect to that of pure V$_2$O$_3$ (2 electrons $per$ atom) in its metallic phase. Both effects then qualitatively suggest a less drastic metal-insulator transition at T$_N$, with an antiferromagnetic phase having both a small gap and a strongly renormalized N\'eel temperature. This is also confirmed by the reduction of the jump of resistivity at T$_N$ in the Ti-doped sample. However, we cannot exclude the presence of an electronic phase separation, although the well evident metallic aspect of the optical conductivity for T$>$T$_N$ seems to indicate, at variance with Cr-doped 0.01 crystal, that metallic domains are predominant over the insulating ones. This calls for further SPEM measurements to investigate this interesting point.

Cr substitution also plays a double role. The incorporation of Cr into V$_2$O$_3$ results in an expansion of the corundum lattice \cite{McWhan-73}. In particular, the expansion determines an increase of the crystal field splitting and a reduction of both the hopping $a_{1g}$-$a_{1g}$ integral and the  $a_{1g}$-$e^{\pi}_{g}$ hybridization energy. These changes in the LDA parameters \cite{Grieger-14} have been suggested to be significant to drive the metal-insulator transition as explained in Ref. \onlinecite{Poteryaev-07}. Moreover, given the $3d^{5}$$4s^{1}$ Cr electronic state, this corresponds to an electron doping with respect to the half-filling configuration of pure V$_2$O$_3$. Both effects correspond to a weak, scarcely Cr-dependent, insulating PI phase (see Fig. 2), with a small insulating gap of about 1000 cm$^{-1}$. Starting from this weak insulating state, the AF-monoclinic phase transition further gaps the optical conductivity but, at variance with pure V$_2$O$_3$, where the half-filling condition favors a large gap, here a smaller gap opens in the AFI phase with an absorption maximum at about 8000  cm$^{-1}$.

\section*{IV. CONCLUSIONS}
In this paper we have discussed the optical properties of the archetypal correlated material V$_2$O$_3$ in all regions of its phase diagram. From a Drude-Lorentz fit we find that titanium acts like an electronic acceptor putting x(Ti) holes in the crystal, and thus reducing the metallicity of the sample. Chromium doping instead induces a phase coexistence with percolative insulating domains, thus drastically lowering the dc conductivity, compared to the pure material.  We have also drawn important information from the high temperature crossover regime, where a depletion in the optical conductivity is formed and the sample shows a ``poor metal'' behavior. We have estimated the degree of correlation by utilizing the temperature dependence of the spectral weight. This method gives results in good agreement with previously proposed estimations providing vanadium sesquioxide to be classified as a very strongly correlated material. Furthermore we discuss for the first time the antiferromagnetic insulating gap magnitude, and we compare it among the various doped samples. The Cr-doped compound shows half the gap of the pure compound due to lattice parameters modification and to the actual electron doping strongly modifying the band structure. This study thus provides a new comprehensive scenario of the Mott-Hubbard physics in V$_2$O$_3$.

\end{document}